\documentclass{iaus}
\usepackage{graphicx}

\def\specchar#1{{\sc #1}}
\def\SrI{\mbox{Sr\,\specchar{i}}}
\def\OI{\mbox{O\,\specchar{i}}}
\def\TiI{\mbox{Ti\,\specchar{i}}}
\def\FeI{\mbox{Fe\,\specchar{i}}}
\def\SiI{\mbox{Si\,\specchar{i}}}
\def\BaII{\mbox{Ba\,\specchar{ii}}}

\title[Spectropolarimetric diagnostics of unresolved magnetic fields] 
{Spectropolarimetric diagnostics of unresolved magnetic fields in the quiet solar photosphere}

\author[Nataliya G. Shchukina \& Javier Trujillo Bueno]   
{Nataliya G. Shchukina$^1$
 \and Javier Trujillo Bueno$^2$}

\affiliation{$^1$Main Astronomical Observatory, National Academy of Sciences,
\\ 27 Zabolotnogo street,
Kiev 03680, Ukraine \\ email: {\tt shchukin@mao.kiev.ua} \\[\affilskip]
$^2$Instituto de Astrof\'{\i}sica de Canarias, 38205 La Laguna, Tenerife, Spain
\\email: {\tt jtb@iac.es}}

\pubyear{2012}
\volume{294}  
\pagerange{1--13}
 \jname{Solar and Astrophysical Dynamos}
\editors{A.G. Kosovichev, E.M. de Gouveia Dal Pino \& Y.Yan.}
\begin{document}

\maketitle

\begin{abstract}
A few years before the Hinode space telescope was launched,
an investigation based on the Hanle effect in atomic and molecular lines indicated
that the bulk of the quiet solar photosphere is significantly magnetized,
due to the ubiquitous presence of an unresolved magnetic field
with an average strength $\langle B \rangle {\approx} 130$ G.
It was pointed out also that this ``hidden" field must be much stronger
in the intergranular regions of solar surface convection than in the granular regions,
and it was suggested that this unresolved magnetic field could perhaps provide
the clue for understanding how the outer solar atmosphere is energized.
In fact, the ensuing magnetic energy density is so significant
that the energy flux estimated using the typical value of 1 km/s
for the convective velocity (thinking in rising magnetic loops)
or the Alfv\'{e}n speed (thinking in Alfv\'{e}n waves generated by magnetic reconnection)
turns out to be substantially larger than that required to balance the chromospheric energy losses.
Here we present a brief review of the research 
that led to such conclusions, with emphasis on a new   
three-dimensional radiative transfer
investigation aimed at determining the magnetization
of the quiet Sun photosphere from the Hanle effect
in the \SrI\ 4607 \AA\ line and the Zeeman effect in \FeI\  lines.
%
\end{abstract}


\firstsection 
\section{Introduction}

The subject of this paper is small-scale (SS) magnetic fields in the quiet solar photosphere.
Most part of the solar surface is covered by the inter-network regions of the quiet Sun,  
which appear empty (i.e., devoid of magnetic signatures) in low resolution magnetograms. However, such regions are magnetized, as shown by the weak Stokes $V$ signals (produced by the longitudinal Zeeman effect) that \cite[Livingston \& Harvey (1971)]{Livingston_Harvey1971} could detect in the interiors of supergranular cells. Over the last 10 years, the complexity of the SS magnetic fields of the quiet solar photosphere has been studied vigorously through high-spatial resolution observations of the polarization produced by the Zeeman effect in \FeI\ lines, using both ground-based telescopes  (e.g.,
\cite[Dom\'{\i}nguez Cerde\~{n}a et al. 2003]{Dominguez_Cerdena_etal2003};
\cite[Khomenko et al. 2003]{Khomenko_etal2003};
\cite[Lites \& Socas-Navarro 2004]{Lites_Socas-Navarro2004};
\cite[Dom\'{\i}nguez Cerde\~{n}a et al. 2006]{Dominguez_Cerdena_etal2006};
\cite[Harvey et al. 2007]{Harvey_etal2007};
\cite[Mart\'{\i}nez Gonz\'{a}lez et al. 2007]{Martinez_Gonzalez_etal2007}),
the HINODE space telescope (e.g.,
\cite[Centeno et al. 2007]{Centeno_etal2007};
\cite[Orozco Su\'{a}rez et al. 2007]{Orozco_Suarez_etal2007};
\cite[Rezaei et al. 2007]{Rezaei_etal2007};
\cite[Lites et al. 2008]{Lites_etal2008};
\cite[Ishikawa \&  Tsuneta 2009]{Ishikawa_Tsuneta2009};
\cite[Asensio Ramos 2009]{Asensio2009}; 
\cite[Mart\'{\i}nez Gonz\'{a}lez et al. 2010]{Martinez_Gonzalez_etal2010}; 
\cite[Danilovic et al. 2010]{Danilovic_etal2010}; 
\cite[Mart\'{\i}nez Pillet et al. 2011a]{Martinez_Pillet_etal2011a};
\cite[Viticchi\'e \& S\'anchez Almeida 2011]{Viticchie-Sanchez_Almeida2011};
\cite[Manso Sainz et al. 2011]{Manso_Sainz_etal2011}), 
and the balloon-borne telescope SUNRISE (e.g., 
\cite[Solanki et al. 2010]{Solanki_etal2010};
\cite[Borrero et al. 2010]{Borrero_etal2010};
\cite[Lagg et al. 2010]{Lagg_etal2010};
\cite[Mart\'\i nez Gonz\'alez et al. 2012]{Martinez_etal2012}).

Diagnostic tools based on the polarization of the Zeeman effect are 
however practically blind to the presence of magnetic fields 
that are randomly oriented on scales too small to be resolved.
Therefore, non-detection of Zeeman polarization does not necessarily 
imply absence of magnetic fields. 

A key question is: how significant is the degree of magnetization of the 
plasma of the quiet Sun ? To answer this question it is necessary to investigate 
how much magnetic flux and energy reside at small (unresolved) scales. 
To explore this ``hidden" magnetism of the quiet Sun 
one has to apply diagnostic techniques based 
on the Hanle effect in atomic and molecular lines, 
ideally complementing them with those based on the Zeeman effect 
(see the review by 
\cite[Trujillo Bueno et al. 2006]{Trujillo_etal2006}).
In this paper we provide a brief summary 
of the research carried out over the last few years, 
with emphasis on three-dimensional radiative transfer investigations 
of the Hanle effect in the \SrI\ line at 4607 \AA\ and of the Zeeman effect in some \FeI\ lines.
%


\section{Polarization in Spectral Lines}

The most powerful tool for the diagnostics of magnetic fields in the 
atmospheres of the Sun and of other stars 
is the interpretation of spectropolarimetric observations 
(e.g., \cite[Landi Degl'Innocenti \& Landolfi 2004]{Landi_DeglInnocenti_Landolfi2004}).
Polarization in spectral lines can be
induced and modified
by several physical mechanisms.
In the quiet Sun the most important
ones are
{\it the Zeeman effect} 
(\cite[Zeeman 1897]{Zeeman1897}),
{\it anisotropic radiation pumping}
(\cite[Kastler 1950]{Kastler1950}),
and
{\it the Hanle effect}
(\cite[Hanle 1924]{Hanle1924}).
%

{\underline{\it The Zeeman effect}}
requires the presence of a magnetic field, which causes the atomic
and molecular energy levels to split into different magnetic sublevels characterized by their
magnetic quantum number $M$. As a result, the wavelength positions of the
$\pi$
($\Delta M = M_u - M_l = 0$),
$\sigma_{\rm blue}$ ($\Delta M = +1$) and
$\sigma_{\rm red}$ ($\Delta M = -1$) transitions do not coincide and,
therefore, their respective polarization signals do not cancel out.
The Zeeman effect is most
sensitive in circular polarization
(quantified by the Stokes $V$ parameter)
with a $V/I$ amplitude 
that
for not too strong fields scales with the ratio, $R$, between the Zeeman splitting
$\Delta \lambda_{B}$
and the
Doppler line-width 
$\Delta \lambda_{D}$:
\begin{equation}
{{V/I} \sim {R}} ={{\Delta \lambda_{B}}\over{\Delta \lambda_{D}}}
\sim {{\lambda B}\over {\surd (T/\alpha)}} \ll {1} ,
\end{equation}
%
with $\alpha$ the atomic weight of the atom under consideration. 
The Stokes
$V$ signal 
changes its sign for opposite orientations of the magnetic field vector.
This so-called {\it longitudinal Zeeman effect} responds to the line-of-sight component of
the magnetic field ($B_{\parallel}$).
On the contrary, {\it the transverse Zeeman effect} responds to the component of
the magnetic field perpendicular to the line of sight ($B_{\perp}$),
producing instead linear polarization
signals quantified by the Stokes $Q$ and $U$ parameters.
The Zeeman polarization $Q/I$ and $U/I$ amplitudes
are approximately proportional to $R^{2}$:
%
\begin{equation}
{{Q/I}\ \ \&\ \ {U/I} \sim {R}^2}.
\end{equation}
%
Taking into account that in the quiet Sun $R{\ll}1$,  
the Stokes $Q/I$ and $U/I$ signals of solar spectral lines are
normally negligible
for intrinsically weak  fields (typically
$B \lesssim 100 $~G).
As a result, Zeeman linear polarization turns out to be
often
below the noise
level of present observational possibilities. It is also important to note that 
if only a fraction, $f$, of the observational resolution element is filled with magnetic field, then 
$B_{\parallel}$ (inferred from $V$) scales with $f$, while $B_{\perp}$ 
(inferred from $Q$ and $U$) scales with $\sqrt{f}$ (cf., Landi Degl'Innocenti \& Landolfi 2004). 
%

{\underline{\it Anisotropic radiation pumping}}.
The spectral line polarization that is induced by scattering processes
in stellar atmospheres
is directly related with
the anisotropic illumination of the atoms.
The presence of a magnetic field is not necessary for producing such a polarization.
The absorption of anisotropic radiation 
produces atomic level polarization
(i. e., 
population imbalances and/or quantum coherence between the
magnetic sublevels pertaining to the upper and/or lower level of
the line transition under consideration), in such a way that the
populations of substates with different values of $\mid{M}\mid$
are different.
Under such circumstances,
the polarization signals of the $\pi$ and $\sigma$ transitions do not cancel out,
even in the
absence of a magnetic field,
simply because the population imbalances among the magnetic
sublevels imply more or less $\pi$ transitions, per unit volume and time,
than $\sigma$ transitions.
Such anisotropic radiation pumping processes
are particularly efficient in creating atomic polarization
if the depolarizing rates caused by elastic collisions are sufficiently low.

{\underline{\it The Hanle effect}}
is the modification of the atomic level polarization
and of its ensuing observable effects on the emergent Stokes $Q$ and $U$ profiles.
It is caused by the action of a magnetic field such that
the corresponding Zeeman splitting is comparable to the inverse lifetime, 
$t_{\rm life}$, 
of the degenerate atomic level under consideration 
(the upper or lower level of the chosen line transition).
For the Hanle effect to operate, the magnetic field vector has to be inclined
with respect to the symmetry axis of the pumping radiation field.
Typically (but not always), the changes in the Stokes $Q$ and $U$ profiles
due to the Hanle effect
consist in
a net depolarization and a rotation of the direction of linear polarization.
If
the azimuth of the magnetic field is uniformly distributed within
the resolution element of the observation, rotations of the direction of linear polarization 
cancel out, but the reduction of the scattering polarization
amplitude with respect to the zero-field case remains.
Therefore, the Hanle effect has the diagnostic potential for
detecting tangled magnetic fields on
subresolution scales in the solar atmosphere.
Fortunately, scattering processes in the solar atmosphere produce a rich linearly-polarized spectrum
(\cite[Stenflo \&  Keller 1997]{Stenflo-Keller1997}).
%
%


\section{Diagnostics of Zeeman-Effect Polarization}

Most of our present empirical knowledge on 
solar surface magnetism
stems from the analysis of the spectral line polarization caused by the  Zeeman effect.
Diagnostic tools based on the Zeeman effect aim at determining
the magnetic field strength,
the magnetic flux, and
the inclination of the magnetic field.
The advantages and disadvantages of 
Stokes diagnostics based on the Zeeman effect 
have been repeatedly discussed in the literature.
An attractive feature is
the relative simplicity of the physics of the Zeeman effect.
The mere detection of the signature of the Zeeman effect in the observed spectral line polarization  
can be directly interpreted as the presence of a magnetic field.
Another good news for the polarization of the 
Zeeman effect as a diagnostic tool 
is the large number of
methods developed
to extract quantitative properties of
the magnetic field 
(see, for example, 
the reviews by
\cite[Solanki 1993]{Solanki1993}
and 
\cite[Khomenko 2006]{Khomenko2006}).
We can mention  
the magnetic line ratio technique 
(see 
\cite[Stenflo 1973]{Stenflo1973};
\cite[Solanki et al. 1992]{Solanki_etal1992}; 
and references therein),
methods based on spectral synthesis in atmospheric models resulting from 
magnetoconvection simulations (see, for example, 
\cite[Khomenko et al. 2005]{Khomenko_etal2005}; 
\cite[Shelyag et al. 2007]{Shelyag_etal2007};
\cite[Danilovic et al. 2010]{Danilovic_etal2010}, 
and references therein),
methods based on exploiting the hyperfine structure of some atoms 
(L\'{o}pez Ariste et al. 2002; Asensio Ramos et al. 2007),
and various inversion methods
(\cite[Skumanich \& Lites  1987]{Skumanich_Lites1987};
\cite[Ruiz Cobo \& del Toro Iniesta 1992]{Ruiz_Cobo_Toro_Iniesta1992};
\cite[S\'{a}nchez Almeida 1997]{Sanchez_Almeida1997};
\cite[Socas-Navarro et al. 2000]{Socas-Navarro_etal2000}).
%

Among the major shortcomings we should mention the following ones.
First, the Zeeman polarization signals
observed in the quiet Sun are
very weak compared to those observed in active regions, so that 
quantitative diagnostics of the
magnetic field properties are always challenged by the presence of
the measurement noise (e.g., Asensio Ramos 2009).
Second, the polarization of the Zeeman effect as a diagnostic tool is practically
{\it blind}
to magnetic fields that are randomly oriented on
scales too small to be resolved.
In other words,
the Zeeman polarization signals tend to cancel out when averaging.
This is exactly the situation in the internetwork regions of the quiet Sun,
where the observed Stokes $V$ signals show an intermittent pattern of positive
and negative polarities at spatial scales as small as the diffraction limit of the telescope used.
For this reason, one expects that Zeeman-polarization inferences of 
$ \langle {\mid} B_{\parallel} {\mid} \rangle $ (the mean unsigned ``vertical" magnetic flux density) 
depend on the spatial resolution of the observation. As it can be seen in figure 3 of \cite[S\'{a}nchez Almeida  \&  Mart\'{\i}nez Gonz\'{a}lez (2011)]{Sanchez_Almeida_Marinez_Gonzalez2011}, such  unsigned flux density seems indeed to be larger at the close to 0.3 arcsec resolution of HINODE ($ \langle {\mid} B_{\parallel} {\mid} \rangle \, {\approx} \, 10 {\rm G}$) than at the 1 arcsec resolution of some ground-based observations ($ \langle {\mid} B_{\parallel} {\mid} \rangle \, {\approx} \, 3 {\rm G}$). 
Such a figure
suggests that the average unsigned vertical magnetic
flux density 
increases with increasing angular
resolution (i.e., with decreasing size of the resolution element $L$), 
as expected for the case of polarization signals
produced by the random association of equal independent structures with
size $l$ smaller than $L$ (see \cite[S\'{a}nchez Almeida 2009]{Sanchez_Almeida2009}). 
Such a tentative extrapolation predicts a magnetic flux density of 36 G or 181 G, depending
on whether the intrinsic size $l$ is 100 km or 20 km,
respectively. 

Unfortunately, there is a large scatter in the unsigned flux density values
obtained from high spatial resolution Zeeman-polarization measurements
taken with the HINODE satellite (spatial resolution of about 0.3 arcsec)  
and the IMAX instrument on board
SUNRISE (spatial resolution of about 0.15 arcsec).  
The values reported 
so far do not seem to show the above-mentioned expected behavior  
(see 
\cite[Solanki et al. 2010]{Solanki_etal2010}; 
\cite[Mart\'{\i}nez Pillet et al. 2011b]{Martinez_Pillet_etal2011b};
\cite[S\'{a}nchez Almeida  \&  Mart\'{\i}nez Gonz\'{a}lez 2011]{Sanchez_Almeida_Marinez_Gonzalez2011};
\cite[Orozco Su\'{a}rez \& Bellot Rubio 2012]{Orozco_Bellot2012}, 
and more references therein).
\cite[S\'{a}nchez Almeida  \&  Mart\'{\i}nez Gonz\'{a}lez (2011)]{Sanchez_Almeida_Marinez_Gonzalez2011}
argue that these apparent inconsistencies are mostly due to 
unaccounted biases in the applied diagnostic techniques.

Summarizing, 
we can conclude
that using only the Zeeman polarization technique 
is not a very suitable strategy 
for investigating magnetic fields that have complex
unresolved geometries.
There is some evidence that the complexity of the SS fields of the quiet Sun 
is so considerable that with the present instrumentation  
Zeeman effect
diagnostics allow us to detect only the tip of the iceberg of the quiet Sun magnetism.
For example,
a recent estimation by 
Pietarila-Graham et al. (2009) predicts that 
$\sim 80$~\%
of the Stokes-$V$ signals existing in the surface dynamo simulations of 
\cite[V\"{o}gler \& Sch\"{u}ssler (2007)]{Vogler_Schussler2007}
would not be observable at the HINODE resolution of 0.3
arcsec (cf., Emonet \& Cattaneo 2001; S\'anchez Almeida et al. 2003).
Fortunately, unresolved magnetic
fields can be detected through the Hanle effect.



\section{Diagnostics of Hanle-Effect Polarization}

Since opposite polarity fields contribute with the same
sign to the Hanle depolarization, 
the Hanle effect in suitably chosen spectral lines was correctly considered 
a potentially powerful tool to explore
SS magnetic fields
that are tangled on scales too small to be resolved (Stenflo 1982).
Another important advantage of the Hanle effect as a diagnostic tool of the quiet Sun 
magnetism is that it is
sensitive to much weaker magnetic fields than those that can be detected through 
the Zeeman effect 
(i.e., in the range from $10^{-3}$ G up
to a few hundred gauss, depending on the line transition).
The main disadvantage is that magnetic  
fields stronger than the Hanle saturation limit
remain virtually indistinguishable
(e.g., in the case of the \SrI\ 4607 \AA\ line, for 
magnetic strengths   
$B > 250$ G). Therefore, both plasma diagnostic tools are rather complementary. 

A useful formula to estimate the magnetic field strength, 
$B_{\rm c}$
(measured in G), for the onset of the Hanle effect in a line transition 
without atomic polarization in its lower level (such as that of Sr {\sc i} at 4607 \AA) is 
(e.g., 
\cite[Trujillo Bueno \& Manso Sainz 1999]{Trujillo-Manso1999})
%
%
\begin{equation}
B_{\rm c}=(1+\delta)B_{\rm H},
\end{equation}
where $\delta={\rm D}^{(2)}/A_{ul}$ is the collisional depolarizing rate (typically due to 
elastic collisions with neutral hydrogen atoms) in units of the Einstein $A_{ul}$ coefficient for 
spontaneous emission, and 
\begin{equation}
{B_{\rm H}} = {1.137 \times 10^{-7}}\,\,{{A_{ul}}\over {g_{\rm L}}}
\end{equation}
is the magnetic strength for which the Zeeman splitting of the line's upper level 
is equal to its natural width ($g_{\rm L}$ being the level's Land\'{e} factor).

A first rough
estimate of the Hanle-effect depolarization
needed to explain some line scattering polarization observations
of quiet regions of the solar atmosphere had suggested 
a tentative lower limit to the mean field strength 
$\langle B \rangle$ of 10 G
(\cite[Stenflo 1982]{Stenflo1982}).
In recent years the Hanle effect has changed from being considered
an exotic theoretical (de)polarization mechanism
to a powerful tool for the diagnostics of the solar magnetism.
Recent advances in the Hanle-effect polarization diagnostics 
of the magnetism of the quiet solar photosphere  
are based on spectral lines like 
the molecular lines of $\rm {C_2}$, MgH and CN
(see 
\cite[Trujillo Bueno et al. 2004]{Trujillo_Bueno_etal2004};
\cite[Asensio Ramos \& Trujillo Bueno 2005]{Asensio-Trujillo2005};
\cite[Shapiro et al. 2011]{Shapiro_etal2011},
and references therein),
the lines of multiplet 
$a\rm{{^5}F} - y\rm{{^5}F^{o}}$
of \TiI\
(\cite[Manso Sainz et al. 2004]{Manso_Sainz_etal2004}; 
\cite[Shchukina \& Trujillo Bueno 2009]{Shchukina_Trujillo2009}), 
and
the \SrI\ 4607 \AA\ line
(\cite[Faurobert-Scholl et al. 1995]{Faurobert-Scholl_etal1995};
\cite[Faurobert et al. 2001]{Faurobert_etal2001};
\cite[Shchukina  \&  Trujillo Bueno 2003]{Shchukina_Trujillo2003}; 
\cite[Trujillo Bueno et al. 2004]{Trujillo_Bueno_etal2004};
\cite[Bommier et al. 2005]{Bommier_etal2005};
\cite[Trujillo Bueno \& Shchukina 2007]{Trujillo_Shchukina2007};
\cite[Shchukina \& Trujillo Bueno 2011]{Shchukina_Trujillo2011}).
The \SrI\  line at 4607 \AA\ 
is particularly convenient for the investigating
the mean magnetization of the ``quiet'' solar photosphere
thanks to  
its conspicuous scattering polarization signals 
(spatially and temporally averaged $\langle Q \rangle/\langle I \rangle \sim 1$~\% near the solar limb)
and its high sensitivity to the Hanle effect (between 10 and 250 G, approximately).
In the next section
we highlight some significant results 
obtained through the application of the Hanle effect 
in the \SrI\ 4607 \AA\ line.


\section{The Hanle Effect in the \SrI\ 4607 \AA\ Line}

Hanle-effect diagnostics using this line relies on a comparison
between the observed scattering polarization $Q/I$ signals 
and those corresponding to the zero-field reference case. The determination 
of the zero-field polarization amplitudes requires theoretical modeling. 
The quantum theory of spectral line polarization described in the monograph by 
Landi Degl'Innocenti \& Landolfi (2004) is a very suitable theoretical framework for 
understanding the polarization produced  
in solar spectral lines for which correlations between the 
frequencies of the incoming and outgoing photons in the scattering events can be neglected, 
like that of \SrI\ at 4607 \AA.
In this theory, the excitation state of the atomic (or molecular) system
is described by the {\it diagonal} and {\it non-diagonal} elements of 
the atomic density matrix corresponding to each atomic level of total angular momentum $J$,
which quantify the level's overall population,  
the {\it population imbalances} between its sublevels, 
and the {\it quantum coherence}
between each pair of them.
Similarly,
the symmetry properties of the radiation field are described by 9 tensors 
which quantify the mean intensity, 
$J^{0}_{0}$,
the degree of anisotropy 
${\cal A}=J^{2}_{0}/J^{0}_{0}$,
and
the breaking of the axial symmetry of the incident radiation field 
(e.g., due to horizontal atmospheric inhomogeneities). 


A suitable model for estimating the Hanle depolarization 
in the \SrI\ 4607 \AA\ line 
is that of a microturbulent field (i.e., that the ``hidden'' field has
an isotropic distribution of orientations within a photospheric volume given
by ${\cal L}^3$, with ${\cal L}$ the mean free path of the line-center photons).
With this assumption for the topology of the ``hidden" field and the two-level atom 
model (see Trujillo Bueno \& Manso Sainz 1999) it is easy to obtain the following 
Eddington-Barbier approximation for estimating the line-center scattering 
polarization amplitude of the emergent radiation in the \SrI\ 4607 \AA\ resonance line 
(see \cite[Shchukina  \&  Trujillo Bueno 2003]{Shchukina_Trujillo2003}):
%
\begin{equation}
\label{SLQ:SLQ_TL}
 {Q} / {I} \approx\,
{{3}\over{2\sqrt{2}}}\,(1-\mu^2)\,{{\cal H}\over{1+{\delta}}}\,{\cal A}\,.
\end{equation}
%
In this approximate expression 
$\mu=\cos\theta$
(with the angle $\theta$ between the solar radius vector through the
observed point and the line of sight),
and
${\cal H}$ represents the Hanle depolarization factor of the microturbulent
field:
\begin{equation}
\label{H}
 {\cal H} = 1/5 \times \{1 + 2/(1+{\Gamma}^{2}) +
2/(1+4\,{\Gamma}^{2})\},
\end{equation}
where
\begin{equation}
\label{gammaL}
 \Gamma \approx {\gamma\,\times\,}{{1}\over{1+{\delta}}}
\end{equation}
and
\begin{equation}
\label{gammaS}
 \gamma=8.79{\times}10^{6}\,B\,g_{\rm L}/A_{ul}.
\end{equation}
The Hanle depolarization factor ${\cal H}=1$ if the magnetic field strength $B=0$ G, 
while ${\cal H}=1/5$ for $B > 250$~G (i.e., ${\cal H}=1/5$ 
for magnetic strengths larger than the Hanle saturation 
field of the \SrI\ 4607 \AA\ line). 

The approximate formula~(\ref{SLQ:SLQ_TL}) shows clearly 
that a reliable estimation of the strength of the ``hidden" photospheric field requires 
adopting realistic values for the collisional depolarizing rate $\delta$,  
and determining the anisotropy ${\cal A}$ of the spectral line radiation 
through radiative transfer calculations in realistic atmospheric models.

{\underline{\it One-dimensional modeling}}.
The first radiative transfer modeling of the scattering polarization 
observed in the \SrI\ 4607 \AA\ line were one-dimensional, using 
a plane-parallel, static semi-empirical model of the quiet solar atmosphere 
and the above-mentioned microturbulent field model
(\cite[Faurobert-Scholl et al. 1995]{Faurobert-Scholl_etal1995}; 
\cite[Faurobert et al. 2001]{Faurobert_etal2001}).
These authors concluded that the mean field strength around a height 
$h\,{\approx}\,300$ km above the 
visible solar surface is between 10 and 25 gauss, assuming the 
single-valued microturbulent field model.  
Given that the atmospheric model used was one-dimensional and static, 
such modeling had to make use of the free parameters of stellar spectroscopy 
(that is, micro and macroturbulent velocities for the line broadening). As shown 
by
\cite[Shchukina  \&  Trujillo Bueno (2003)]{Shchukina_Trujillo2003} 
the choice made by Faurobert et al. for such free parameters
yields artificially low values for the mean
strength of the hidden field. 
This is because the
scattering polarization amplitudes calculated
by
those authors
for the zero-field reference case,
$(Q/I)_{B=0}$,
was
seriously underestimated.

%
{\underline{\it Three-dimensional modeling using a hydrodynamical (HD) model}}.
The plasma
of the solar atmosphere is inhomogeneous and dynamic, very different from a uniform static plane-parallel configuration. 
In order to improve the reliability of diagnostic tools based on the Hanle
effect,
\cite[Shchukina  \&  Trujillo Bueno (2003)]{Shchukina_Trujillo2003}, 
\cite[Trujillo Bueno et al. (2004)]{Trujillo_Bueno_etal2004}
and
\cite[Trujillo Bueno \&  Shchukina (2007)]{Trujillo_Shchukina2007}
developed a new approach based on multilevel, non-LTE scattering polarization
calculations in three-dimensional (3D) models of the solar photosphere. In order to 
achieve a reliable calculation of the anisotropy of the spectral line radiation within the 
solar photosphere, it is very important to use realistic 3D models. 
They 
used a 3D 
photospheric model resulting 
from the hydrodynamical simulations of solar surface convection
by \cite[Asplund et al. (2000)]{Asplund_etal2000}.
The detailed non-LTE studies of  
\FeI\ 
(\cite[Shchukina \& Trujillo Bueno 2001]{Shchukina_Trujillo_Bueno_2001_Fe}),
\OI\ 
(\cite[Shchukina et al. 2005]{Shchukina_Trujillo_Bueno_2005_HD}),
\TiI\ 
(\cite[Shchukina \& Trujillo Bueno 2009]{Shchukina_Trujillo2009}),
\BaII\ 
(\cite[Shchukina et al. 2009]{Shchukina_Olshevsky_Khomenko2009_Ba}),
 \SiI\
(\cite[Shchukina et al. 2012]{Shchukina_etal2012}),
and of the intensity and polarization of the Sun's continuum spectrum
(\cite[Trujillo Bueno \& Shchukina 2009]{Trujillo_Bueno_Shchukina_2009_cont})
show that this 3D HD model provides a suitable representation
of the thermal and dynamical structure of the quietest regions of the Sun's photosphere.

%
%
The 3D radiative transfer approach allowed 
\cite[Trujillo Bueno et al. (2004)]{Trujillo_Bueno_etal2004}
to achieve a reliable calculation of the linear polarization amplitudes that scattering processes
in the solar photosphere would produce in the \SrI\ 4607 \AA\ line if there
were no magnetic field (see also Shchukina \& Trujillo Bueno 2003).  
These authors
confronted low resolution observations of the center-to-limb 
variation of the scattering polarization in the \SrI\ 4607 \AA\ line  
with the $Q/I$ signals that result 
from spatially averaging the emergent $Q$ and $I$
profiles calculated in the above-mentioned 3D HD model.
The Stokes $I$, $Q$ and $Q/I$ profiles of the \SrI\ 4607 \AA\ line that 
they used had been observed  
by several authors, both during a minimum and a maximum of the solar activity cycle (see
references in 
\cite[Trujillo Bueno et al. 2004]{Trujillo_Bueno_etal2004}).
The absence of any clear variation in the line-center amplitudes of the observed $Q/I$ profiles 
suggested that they do not seem to be modulated by the solar cycle.
\cite[Trujillo Bueno et al. (2004)]{Trujillo_Bueno_etal2004}  
found also that the synthetic intensity profiles of the \SrI\ 4607 \AA\ line 
(which    
they 
obtained by taking
fully into account the Doppler shifts of the convective flow velocities in the 3D model)
are in good agreement with the Stokes-$I$ observations when the meteoritic
strontium abundance is chosen.
However, the calculated $Q/I$ line-center amplitudes 
turned out to  be substantially larger than the observed ones.
They used realistic values for the depolarizing rates due to elastic collisions
with neutral hydrogen atoms (see equation 33 of Faurobert-Scholl et al. 1995) and  
concluded that such a significant discrepancy 
between the observed and calculated scattering polarization amplitudes 
indicated the presence of an unresolved, ``hidden" magnetic field on sub-resolution scales. 

In order to estimate the mean strength of such an unresolved magnetic field, 
\cite[Trujillo Bueno et al. (2004)]{Trujillo_Bueno_etal2004} used the microturbulent field model and 
two functional forms for the Probability Distribution Function, PDF($B$), 
describing the fraction of the 3D HD model occupied by magnetic fields of strength $B$. 
For the idealized case of a single-valued field (${\rm PDF}(B) = \delta (B - \langle B \rangle)$) they 
obtained 
$\langle B \rangle \approx 60$ at an atmospheric height of about 300 km, 
and a clear indication that 
$\langle B \rangle$ decreases with height in the quiet solar photosphere.
\footnote{It is of interest to mention that this  
conclusion for the idealized case 
of a single-valued field was later confirmed by Bommier et al. (2005), 
who applied the 1D modeling approach with a more judicious choice than 
Faurobert-Scholl et al. (1995) and Faurobert et al. (2001) for the values of the micro and macroturbulent velocities.} 
For the more realistic case of a 
${\rm PDF}(B) = \left( {1}/{\langle B \rangle}\right ) \cdot \exp \left ( {-B}/ {\langle B \rangle}\right )$  
the result  
was $\langle B \rangle \approx 130$ G at an atmospheric height of about 300 km, again with a clear indication that 
$\langle B \rangle$ is significantly larger in the low photosphere. 
This natural choice for the shape of the PDF (i.e., an exponential shape) 
was supported both by observations
(e.g., 
\cite[Khomenko et al. 2003]{Khomenko_etal2003})
and by numerical experiments of turbulent dynamos and
magnetoconvection
(e.g., 
\cite[Cattaneo 1999]{Cattaneo1999};
\cite[Stein \& Nordlund 2003]{Stein_Nordlund2003}).
\cite[Trujillo Bueno et al. (2004)]{Trujillo_Bueno_etal2004}
concluded that the magnetization of the 
quiet solar photosphere is very significant ($\langle B \rangle \approx 130$ G), 
substantially larger than the tentative lower limit to $\langle B \rangle $ of 10 G given by 
\cite[Stenflo (1982)]{Stenflo1982}, 
or the 10--25 G given by 
\cite[Faurobert-Scholl et al. (1995)]{Faurobert-Scholl_etal1995} 
and
\cite[Faurobert et al. (2001)]{Faurobert_etal2001}.
With $\langle B \rangle \approx 130$ G
the magnetic energy density 
$E_m = {\langle B \rangle}^{2}/{8\pi} \approx 1300$ erg\,cm$^{-3}$, which is a truly 
significant fraction (${\sim}10\%$) of the kinetic energy density produced by convective
motions in the low photosphere of the HD  
model\footnote{This result and the fact that the scattering polarization 
observed in the \SrI\ 4607 \AA\ line does not seem to be
modulated by the solar cycle, 
suggested that a small-scale dynamo associated
with ``turbulent'' motions
within a given convective domain of ionized gas plays
a significant role for producing 
the ``hidden" magnetic fields diagnosed through the Hanle effect.}.
\cite[Trujillo Bueno et al. (2004)]{Trujillo_Bueno_etal2004} 
estimated the ensuing energy flux by using the typical value of 1 km ${\rm s}^{-1}$ for the 
convective velocities (thinking in rising magnetic loops) or for the Alfv\'en speed 
(thinking in MHD waves), and concluded that in the upper solar photosphere 
it is about $10$ times larger than that required to balance the chromospheric energy losses. 
Some recent investigations based on HINODE observations of 
the Zeeman effect in \FeI\ lines 
have supported this conclusion (e.g., Ishikawa \& Tsuneta 2009; Mart\'{i}nez Gonz\'alez et al. 2010). 

The above-mentioned result (obtained from the Hanle effect in the \SrI\ 4607 \AA\ line)  
may be summarized by saying that at a height of about 300 km above the 
visible solar surface $\langle B \rangle \approx 130$ G, 
when no distinction is made between the granular and intergranular regions. 
A second conclusion could be obtained through a joint analysis of the Hanle effect 
in C$_2$ lines and in the \SrI\ 4607 \AA\ line. 
Trujillo Bueno et al. (2004)
(see also Trujillo Bueno et al. 2006)
found that while the plasma of the upflowing cell centers 
is weakly magnetized (with $\langle B \rangle \sim 10$ G), 
the downward-moving intergranular lane plasma is pervaded 
by relatively strong tangled magnetic fields at sub-resolution scales, 
with $\langle B \rangle$ larger than the Hanle saturation field of the \SrI\ 4607 \AA\ line 
(i.e., with $\langle B \rangle > 250$ G).


\begin{figure}[b]
\begin{center}
 \includegraphics[width=4.5in]{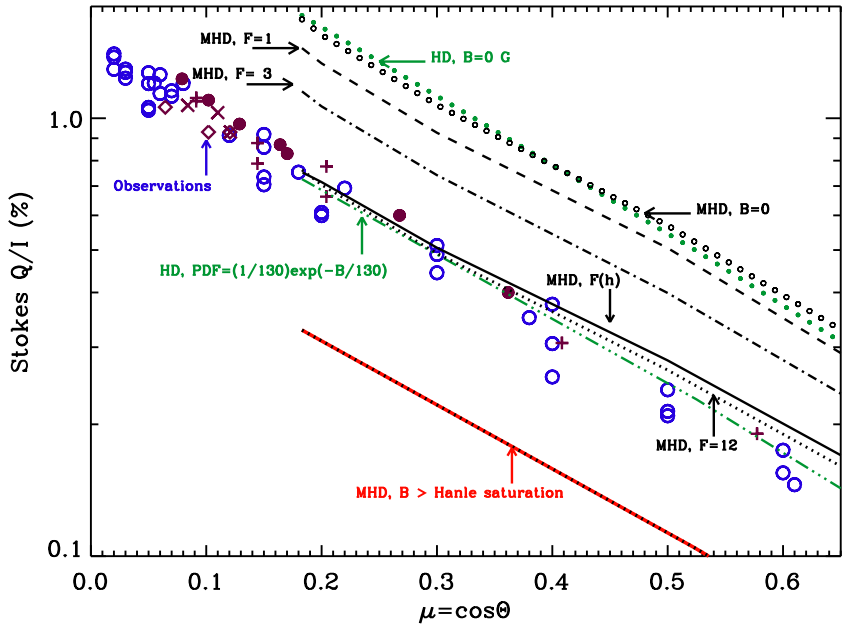} 
 \caption{Center-to-limb variation of the spatially averaged $Q/I$ scattering
amplitudes of the \SrI\ 4607 \AA\ line. The different symbols correspond
to various observations taken by several authors during a minimum and a
maximum of the solar activity cycle 
(see references in 
\cite[Trujillo Bueno et al. 2004]{Trujillo_Bueno_etal2004}).
The two green lines show scattering polarization amplitudes calculated in the
HD model, without including any magnetic field (green filled circles) and
including the Hanle depolarization of a microturbulent field with an exponential
PDF characterized by a mean field strength
$\langle B\rangle = 130$~G
(green dashed-tree-dotted line).
The black lines show $Q/I$ amplitudes calculated in the
MHD model, neglecting its magnetic field (black open circles) and taking into
account the Hanle depolarization of the model's magnetic field (black dashed
line, 
$F=1$). As shown by the black solid line, the observations can be approximately
fitted by multiplying each grid-point magnetic strength by a height-dependent
factor 
$F(h)$
(which implies the height variation of 
$\langle B\rangle$
given by the solid line
of Fig.~2). 
Practically the same result is obtained with a constant
scaling factor 
$F = 12$
(which implies the unrealistic height variation of 
$\langle B\rangle$
given by the dotted line of Fig.~2). 
The red solid line shows the calculated
scattering polarization amplitudes when imposing 
$B > B_{\rm satur} = 250$~G 
at each grid point in the MHD model.}
   \label{fig1}
\end{center}
\end{figure}



\begin{figure}[b]
\begin{center}
 \includegraphics[width=4.5in]{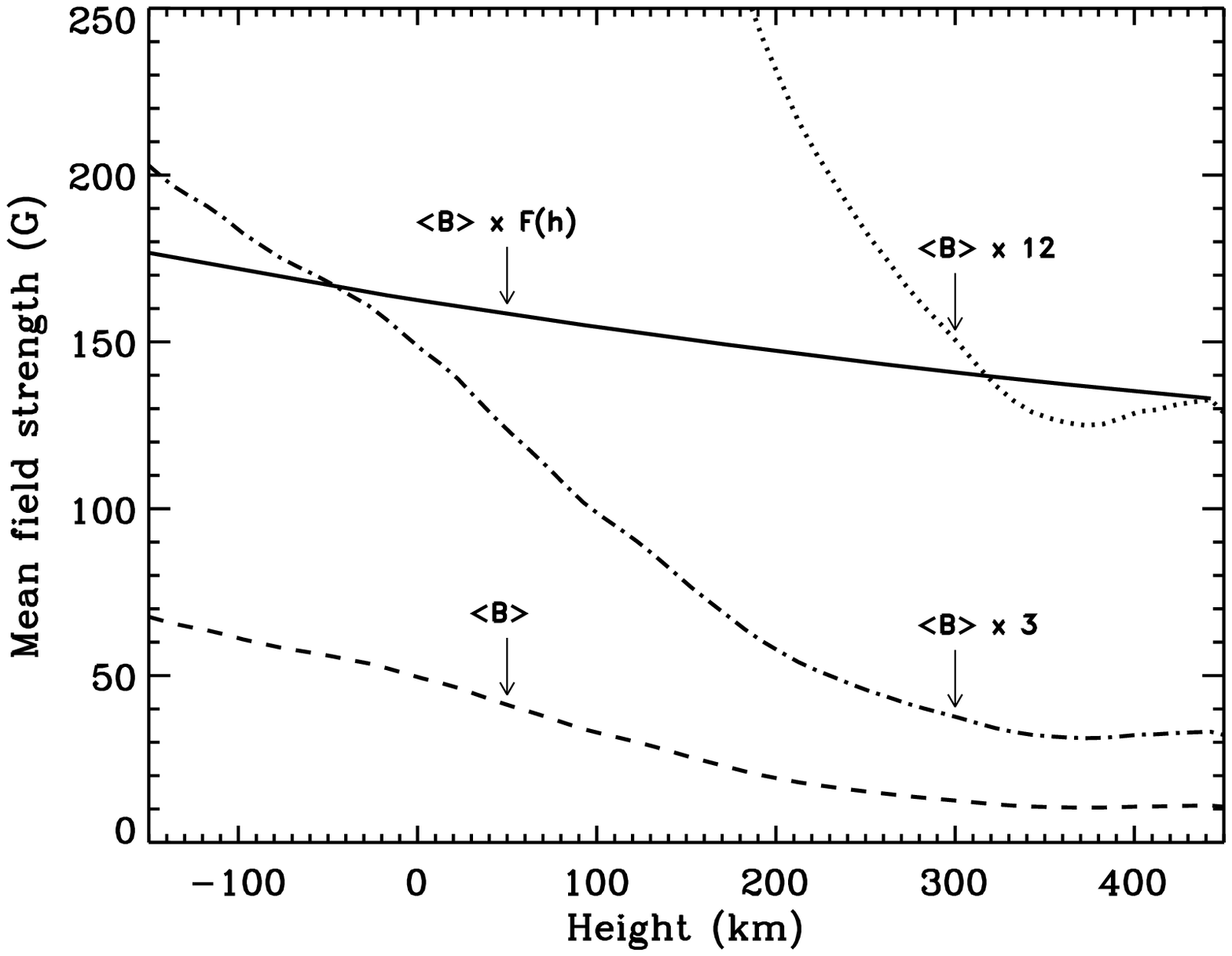} 
 \caption{Height-variation of the mean field strength corresponding to four 
scaling factors $F$ of the magnetic strength of the MHD model. 
Dashed-line: $F = 1$ (i.e.,
as in the original MHD model). 
Dashed-dotted line: 
$F = 3$ (i.e., as suggested by the
Zeeman polarization investigation of 
\cite[Danilovic et al. 2010]{Danilovic_etal2010}).
Dotted line: $F = 12$. 
Solid line: 
a height-dependent scaling factor that explains simultaneously 
the center-to-limb variation of the observed scattering polarization 
$Q/I$
in the \SrI\ 4607 \AA\ line 
and the Zeeman effect polarization in the \FeI\ 6302.5 \AA\ line  
observed with the spectropolarimeter of the Solar
Optical telescope 
onboard of the Hinode satelite
(\cite[Tsuneta et al. 2008]{Tsuneta_etal2008}).
The arrows at $h \approx 60$~km and
$h \approx 300$~km indicate the approximate atmospheric heights around which the observed Zeeman
and Hanle signals are produced, respectively.}
\label{fig2}
\end{center}
\end{figure}


{\underline{\it Three-dimensional modeling using an MHD model
with surface dynamo action. }}

Given that the 3D HD model is unmagnetized Trujillo Bueno et al. (2004) 
had to make the following two 
hypothesis on the unresolved magnetic field that produces Hanle depolarization:
(1) the magnetic field is tangled at scales smaller than the mean free path of the line-center photons,  with an isotropic distribution of directions and (2) the shape of the PDF is exponential. In order to determine the magnetization of the quiet Sun photosphere without using such two approximations,  \cite[Shchukina \& Trujillo Bueno (2011)]{Shchukina_Trujillo2011} have investigated the Hanle effect of the \SrI\ 4607 \AA\ line in a 3D photospheric model resulting from the magneto-convection simulations with surface dynamo action of \cite[V\"{o}gler \& Sch\"{u}ssler (2007)]{Vogler_Schussler2007} (hereafter, the MHD model), which show a complex small-scale magnetic field that results from dynamo amplification of a weak seed field. 

The results of our investigation in the MHD model are summarized in Figs. 1 and 2. 
For the zero-field reference case, the scattering polarization $Q/I$
amplitudes computed in the MHD model (after forcing the magnetic field strength to be zero at each spatial grid point) are similar to those obtained in the HD model
(see Fig.~1, and compare the black open circles with the green filled circles).
The small but noticeable differences between the two results 
are due to the fact that the thermodynamic structure 
of the MHD and HD models are not fully identical.
The black dashed line of Fig. 1 shows that the Hanle depolarization in the \SrI\ 4607 \AA\ line 
produced by the actual magnetic field 
of the MHD model (see dashed line of Fig.~2)
is too small 
to explain the observed Stokes $Q/I$ signals. 
This is
because the average field strength at a height of about 300 km in the MHD model is only
$\langle B \rangle \approx 15$~G 
(i.e., an order of magnitude smaller than the 
$\langle B \rangle$ value inferred by 
\cite[Trujillo Bueno et al. 2004]{Trujillo_Bueno_etal2004}). 
Clearly, the level of small-scale magnetic activity of the MHD surface
dynamo model 
(whose magnetic Reynolds number is rather low, i.e. $R_m \approx 2600$)
is significantly weaker than that of the real quiet Sun photosphere.
The observed
$Q/I$ amplitudes can however be explained after multiplying each grid-point
magnetic strength by a scaling factor $F=12$ (see the dotted line in Figs.~1 and 2), which implies that  
$\langle B \rangle \approx 130$~G in the upper photosphere of the model. 
As shown by the dotted line in Fig.~2, scaling the magnetic field strength of
the MHD model with a constant factor 
$F=12$ 
implies an unrealistic height-variation
of the mean field strength $\langle B \rangle$. 
The resulting large magnetic field strength values 
in the region of formation of the Stokes $V$ Zeeman
signals of the \FeI\ 6302.5 \AA\ line  
(i.e., around 60 km in the MHD model)
would produce synthetic Stokes profiles in contradiction with those observed. 
In fact, $F = 12$ is significantly larger than the tentative scaling
factor $F=3$ needed by 
\cite[Danilovic et al. (2010)]{Danilovic_etal2010}
for explaining the histograms of the polarization
signals produced by the Zeeman effect in the 
\FeI\ lines at 6301.5 \AA\ and 
6302.5 \AA. 
On the
other hand, $F =3$
is too low a value for explaining the $Q/I$ observations
of Fig.~1. 

Therefore, 
in order to explain both
the Hanle depolarization 
of the \SrI\ 4607 \AA\ line 
and
the Zeeman signals  in the \FeI\ lines, 
Shchukina \& Trujillo Bueno (2011) scaled the magnetic strength of the MHD model 
by a height-dependent factor, $F(h)$,  
varying between 
$F \approx 3$ at $h \approx 0$~km
and
$F \approx 12$ at $h \approx 300$~km (see the black solid curves in Figs.~1 and 2).  
With this height-dependent scaling factor, which implies 
$\langle B \rangle \approx 160$~G in the low photosphere and 
$\langle B \rangle \approx 130$~G in the upper photosphere,  
it is possible to explain both  
the scattering polarization observed in  
the \SrI\ 4607 \AA\ line and the Zeeman signals observed with HINODE in the \FeI\ lines. 
Note that with the assumed height-dependent scaling factor 
the scattering polarization amplitudes of the \SrI\ 4607 \AA\ line 
computed in the MHD model are close to the values 
calculated in the HD model 
assuming a microturbulent field
with an exponential PDF characterized by 
$\langle B\rangle = 130$~G (dashed-three-dotted line in Fig.~1).
It is also noteworthy that in the MHD model whose magnetic strength has been scaled with $F(h)$, 
a significant fraction of the model's
granular plasma that contributes to the scattering polarization of the 
$\rm C_2$ lines 
is magnetized  with
$\langle B \rangle \sim 10$~G.
We can thus conclude that the investigation by   
\cite[Shchukina \& Trujillo Bueno (2011)]{Shchukina_Trujillo2011}
reinforces the conclusions of 
\cite[Trujillo Bueno et al. (2004)]{Trujillo_Bueno_etal2004}. 
%


\section{Concluding comments}

Information on our investigations of the Hanle effect 
in 3D models of the quiet solar photosphere can be found in 
\cite[Shchukina \& Trujillo Bueno (2003)]{Shchukina_Trujillo2003}, \cite[Trujillo Bueno et al. (2004)]{Trujillo_Bueno_etal2004}, \cite[Asensio Ramos \& Trujillo Bueno (2005)]{Asensio-Trujillo2005}, \cite[Trujillo Bueno et al. (2006)]{Trujillo_Bueno_etal2006}, 
\cite[Trujillo Bueno \& Shchukina (2007)]{Trujillo_Shchukina2007} and 
\cite[Shchukina \& Trujillo Bueno (2011)]{Shchukina_Trujillo2011}. Here we summarize the main results:

\begin{itemize}
\item
{The quiet solar photosphere is permeated by a small-scale magnetic field, whose
average strength varies approximately 
between $\langle B \rangle \approx 160$ G in the low photosphere ($h\,{\approx}\,60$ km) and 
$\langle B \rangle \approx 130$ G in the upper photosphere ($h\,{\approx}\,300$ km), 
when no distinction is made between granular and intergranular regions.}
\item
{Such a magnetic field is organized at the spatial scales of the solar granulation pattern, 
with relatively weak fields above the granule cell centers and with much stronger fields above the intergranular lanes.
}
\item
{In the upper photosphere, the energy flux estimated 
using the typical value of
1 km\,s$^{-1}$ 
for the convective velocity (thinking in rising magnetic loops) 
or the Alfv\'{e}n speed 
(thinking in MHD waves) turns out to be
an order of magnitude larger
than that required to balance the radiative energy losses from the solar chromosphere.}
\item
{The downward-moving intergranular lane plasma is pervaded by relatively strong tangled
magnetic fields at sub-resolution scales, with $\langle B \rangle > 250$ G. 
This conclusion implies  
that most of the flux and magnetic energy reside on still unresolved scales 
in the intergranular plasma. This leads us to speculate that it is here in 
the turbulent downdrafts where significant ``local" dynamo action is taking place.}
\end{itemize}

{}

\end{document}